# Radioactive contamination of ZnWO$_4$ crystal scintillators

P. Belli[a], R. Bernabei[a,b,1], F. Cappella[c,d], R. Cerulli[e], F.A. Danevich[f],
A.M. Dubovik[g], S. d'Angelo[a,b], E.N. Galashov[h], B.V. Grinyov[g], A. Incicchitti[c,d],
V.V. Kobychev[f], M. Laubenstein[e], L.L. Nagornaya[g], F. Nozzoli[a,b], D.V. Poda[e,f],
R.B. Podviyanuk[f], O.G. Polischuk[f], D. Prosperi[c,d,2], V.N. Shlegel[h], V.I. Tretyak[f],
I.A. Tupitsyna[g], Ya.V. Vasiliev[h], Yu.Ya. Vostretsov[g]

[a] *INFN sezione Roma "Tor Vergata", I-00133 Rome, Italy*
[b] *Dipartimento di Fisica, Università di Roma "Tor Vergata", I-00133 Rome, Italy*
[c] *INFN sezione Roma "La Sapienza", I-00185 Rome, Italy*
[d] *Dipartimento di Fisica, Università di Roma "La Sapienza", 00185 Rome, Italy*
[e] *INFN, Laboratori Nazionali del Gran Sasso, I-67010 Assergi (AQ), Italy*
[f] *Institute for Nuclear Research, MSP 03680 Kyiv, Ukraine*
[g] *Institute for Scintillation Materials, 61001 Kharkiv, Ukraine*
[h] *Nikolaev Institute of Inorganic Chemistry, 630090 Novosibirsk, Russian Federation*

**Abstract**

The radioactive contamination of ZnWO$_4$ crystal scintillators has been measured deep underground at the Gran Sasso National Laboratory (LNGS) of the INFN in Italy with a total exposure 3197 kg × h. Monte Carlo simulation, time-amplitude and pulse-shape analyses of the data have been applied to estimate the radioactive contamination of the ZnWO$_4$ samples. One of the ZnWO$_4$ crystals has also been tested by ultra-low background γ spectrometry. The radioactive contaminations of the ZnWO$_4$ samples do not exceed 0.002 – 0.8 mBq/kg (depending on the radionuclide), the total α activity is in the range: 0.2 – 2 mBq/kg. Particular radioactivity, β active $^{65}$Zn and α active $^{180}$W, has been detected. The effect of the re-crystallization on the radiopurity of the ZnWO$_4$ crystal has been studied. The radioactive contamination of samples of the ceramic details of the set-ups used in the crystals growth has been checked by low background γ spectrometry. A project scheme on further improvement of the radiopurity level of the ZnWO$_4$ crystal scintillators is briefly addressed.

*Keywords:* ZnWO$_4$ crystal; Scintillation detector; Radiopurity, Low background measurement
PACS: 29.40.Mc

## 1. INTRODUCTION

The luminescence of zinc tungstate (ZnWO$_4$) was studied sixty years ago [1]. Large volume ZnWO$_4$ single crystals of comparatively high quality were grown [2] and studied as scintillators in the eighties [3]. The main characteristics of the ZnWO$_4$ scintillators are given in Table 1. The material is non-hygroscopic and chemically resistant. The use of ZnWO$_4$ scintillators was proposed to search for double beta decay in [4] for the first time. The first low background measurement with a small ZnWO$_4$ sample (mass of 4.5 g) was performed in the Solotvina Underground Laboratory (Ukraine) at a depth of ≈ 1000 m of water equivalent (m w.e.) in order to study its radioactive contamination, and to search for double beta decay of zinc and tungsten isotopes [5]; the possibilities to use ZnWO$_4$ crystals in the field of dark matter were also discussed. The luminescence of ZnWO$_4$ down to helium temperature was studied in Ref. [6], and subsequently investigations of ZnWO$_4$ crystals as scintillating bolometers have recently been performed [7, 8, 9].

---

[1] Corresponding author; E-mail address: rita.bernabei@roma2.infn.it
[2] Deceased

Table 1. Properties of the ZnWO$_4$ crystal scintillators.

| Density (g/cm$^3$) | 7.87 [2] |
|---|---|
| Melting point (°C) | 1200 [2] |
| Structural type | Wolframite [10, 11, 12, 13] |
| Cleavage plane | Marked (010) [14] |
| Hardness (Mohs) | 4 – 4.5 [15] |
| Wavelength of emission maximum (nm) | 480 [1, 2,15] |
| Refractive index | 2.1 – 2.2 [15] |
| Effective average decay time$^*$ (μs) | 24 [5] |

* For γ rays, at room temperature.

The radioactive contamination of a 119 g ZnWO$_4$ scintillator was measured to be on the mBq/kg level in the Solotvina Underground Laboratory [16, 9]. Long-time low background scintillation measurements using several ZnWO$_4$ crystal scintillators (with mass in the range 0.1 – 0.7 kg) have been performed at the LNGS with the aim to search for double β processes in zinc and tungsten isotopes [17, 18, 19]. The data collected with different ZnWO$_4$ crystals in the same set-up also allow the estimation of the level of the radioactive contamination of the material. One sample has also been tested by ultra-low background HP Ge γ spectrometry. The effect of the re-crystallization procedure on the radioactive contamination of this material has also been investigated. Moreover, a few samples of ceramics, the most contaminated details in the set-ups used in the crystals growth, have also been measured by a low background HP Ge detector.

## 2. ZnWO$_4$ CRYSTAL SCINTILLATORS

Four clear, slightly colored ZnWO$_4$ crystal scintillators have been used in the present study. All the samples are listed in Table 2. The samples ZWO-1 and ZWO-2 were produced in the Institute for Scintillation Materials (ISMA, Kharkiv, Ukraine) from crystal ingots grown in platinum crucibles by the Czochralski method [16, 20]. The ZnWO$_4$ compounds to grow the crystals were synthesized from two batches of zinc oxide from different producers. The crystal ZWO-3 was obtained by re-crystallization from the sample ZWO-2 at the ISMA. The sample ZWO-4 was produced in the Nikolaev Institute of Inorganic Chemistry (Novosibirsk, Russia) by the low-thermal gradient Czochralski technique also in platinum crucible [21].

## 3. MEASUREMENTS

### 3.1. Low background scintillation measurements

The measurements (see Table 2) have been carried out in the DAMA/R&D set-up [17, 18, 22] at the LNGS having average overburden of about 3600 m w.e. In each measurement the ZnWO$_4$ crystal was fixed inside a cavity of ⌀49 × 59 mm in the central part of a polystyrene light-guide 66 mm in diameter and 312 mm in length. The cavity was filled up with high purity silicone oil. The light-guide was optically connected, on the opposite sides, to two low radioactive EMI9265–B53/FL 3 inches photomultipliers (PMT). The light-guide was wrapped by PTFE reflection tape. The detector was surrounded by Cu bricks and sealed in a low radioactive air-tight Cu box continuously flushed with high purity nitrogen gas (stored deeply underground for a long time) to avoid the presence of residual environmental radon. The Cu box was surrounded by a passive shield made of 10 cm of high purity Cu, 15 cm of low radioactive lead, 1.5 mm of cadmium and 4/10 cm polyethylene/paraffin to reduce the external background. The whole shield has been closed inside a Plexiglas box, also continuously flushed by high purity nitrogen gas.

Table 2. Low background measurements with ZnWO$_4$ crystal scintillators. The times of measurements ($t$), the energy resolutions for the 662 keV γ line of $^{137}$Cs (FWHM), and the background counting rate in different energy intervals are presented. The measurements Run 3 and Run 5 were carried out in the modified set-up with additional quartz light-guides to suppress γ rays from PMTs.

| Run | Crystal | Size Mass Producer | $t$ (h) | FWHM (%) | Background counting rate in counts/(day × keV × kg) in the energy intervals (MeV) | | |
|---|---|---|---|---|---|---|---|
| | | | | | 0.2 – 0.4 | 0.8 – 1.0 | 2.0 – 2.9 |
| 1 | ZWO-1 | 20 × 19 × 40 mm 117 g ISMA [a] | 2906 | 12.6 | 1.71(2) | 0.25(1) | 0.0072(7) |
| 2 | ZWO-2 | ∅44 × 55 mm 699 g ISMA | 2130 | 14.6 | 1.07(1) | 0.149(3) | 0.0072(4) |
| 3 | ZWO-3 | ∅27 × 33 mm 141 g ISMA (re-crystallization of ZWO-2) | 994 | 18.2 | 1.54(4) | 0.208(13) | 0.0049(10) |
| 4 | ZWO-4 | ∅41 × 27 mm 239 g NIIC [b] | 834 | 14.2 | 2.38(4) | 0.464(17) | 0.0112(12) |
| 5 | | | 4305 | 13.3 | 1.06(1) | 0.418(7) | 0.0049(4) |

[a] Institute for Scintillation Materials, Kharkiv, Ukraine;
[b] Nikolaev Institute of Inorganic Chemistry, Novosibirsk, Russia.

In order to suppress the background caused by γ rays from the PMTs, two polished high purity quartz light-guides (∅66 × 100 mm) were optically connected to the opposite sides of the polystyrene light-guide during the Run 3 and Run 5.

An event-by-event data acquisition system accumulates the amplitude and the arrival time of the events. The sum of the signals from the PMTs was recorded with the sampling frequency of 20 MS/s over a time window of 100 μs by a 8 bit transient digitizer (DC270 Acqiris).

The energy scale and the energy resolution of the ZnWO$_4$ detectors were measured by means of $^{22}$Na, $^{60}$Co, $^{133}$Ba, $^{137}$Cs, and $^{228}$Th γ sources. The energy dependence of the energy resolution can be fitted by the function: $FWHM_\gamma(\text{keV}) = \sqrt{a + b \cdot E_\gamma}$, where $E_\gamma$ is the energy of γ quanta in keV. For instance, the energy spectra accumulated by the detector ZWO-4 with $^{60}$Co, $^{137}$Cs and $^{228}$Th γ sources are shown in Fig. 1. The parameters $a$ and $b$ were determined as $a = 2398(570)$ keV$^2$ and $b = 7.96(72)$ keV, respectively. Both the calibration and the background data were taken in the energy interval ~ 0.05 – 4 MeV.

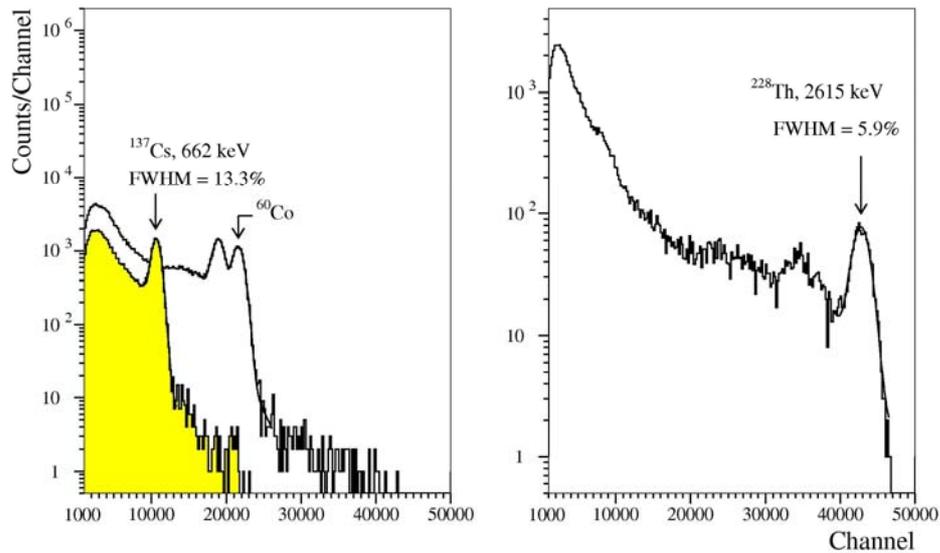

Figure 1. (Color online) Energy distributions measured by ZnWO$_4$ detector ($\varnothing 41 \times 27$ mm) with $^{137}$Cs, $^{60}$Co and $^{228}$Th γ sources.

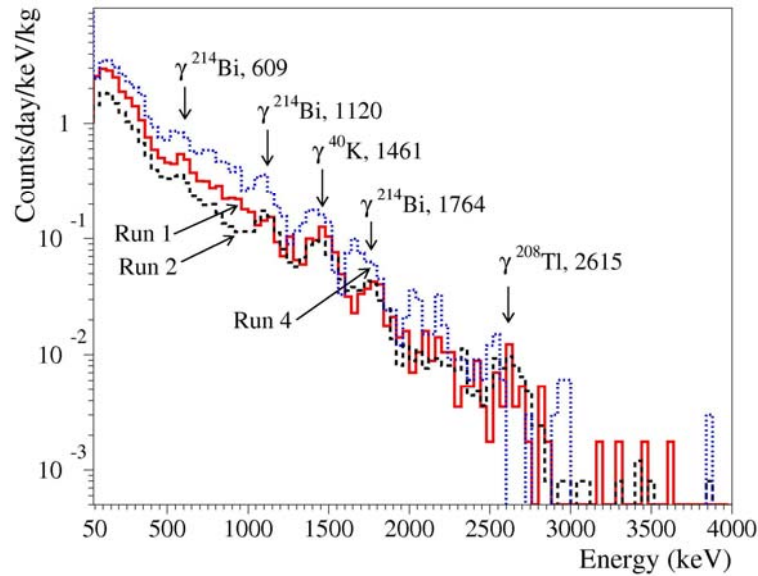

Figure 2. (Color online) Energy distributions of the ZnWO$_4$ scintillators measured in the low background set-up during Runs 1 (solid red line), 2 (dashed black line), and 4 (dotted blue line). Energies of γ lines from residual contaminations are in keV.

The energy distributions accumulated over Runs 1, 2, and 4 with the ZnWO$_4$ scintillation detectors in the low background set-up are shown in Fig. 2. The background spectra accumulated over Runs 2 and 3 with ZnWO$_4$ crystals before and after the re-crystallization are depicted in Fig. 3 (Top) while the energy spectra accumulated over Runs 4 and 5 (before and after the installation of the additional quartz light-guides) are shown in Fig. 3 (Bottom). The spectra are normalized on the mass of the crystals and time of the measurements. A few peaks in the spectra can be ascribed to γ quanta of naturally occurring radionuclides $^{40}$K, $^{214}$Bi ($^{238}$U chain) and $^{208}$Tl ($^{232}$Th) from materials of the set-up. As one can see from Fig. 3 (Bottom), the background spectrum measured over the Run 5 has also a peculiarity: a comparatively wide distribution in the energy interval ≈ 0.6 – 1.1 MeV. Taking into account the α/β ratio[3], this peak is mainly due

---

[3] The relative light yield for α particles as compared with that for γ quanta (β particles) can be expressed through α/β ratio, defined as ratio of α peak position in the energy scale measured with γ sources to the real energy of α

to the radioactive contamination of the crystal ZWO-4 by α active nuclides of $^{232}$Th, $^{235}$U and $^{238}$U families; this statement will further be proved by the pulse-shape discrimination in subsection 3.2.2. The background counting rates of the ZnWO$_4$ detectors in the energy intervals 0.2 – 0.4, 0.8 – 1.0, and 2.0 – 2.9 MeV are given in Table 2.

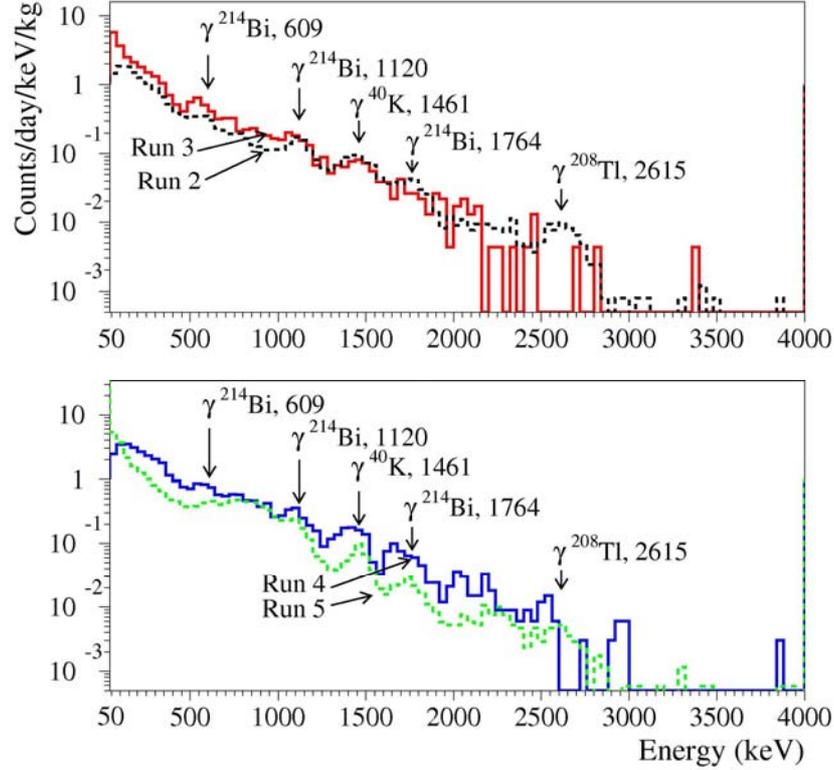

Figure 3. (Color online) Top: Energy distributions of the ZnWO$_4$ scintillators measured in the low background set-up during Run 2 (ZWO-2, dashed black line) and Run 3 (ZWO-3, solid red line). As one can see the re-crystallization did not improve the level of the ZnWO$_4$ detector background. Bottom: The energy spectra of the ZnWO$_4$ crystal scintillator ZWO-4 measured in the low background set-up during Run 4 (solid blue line), and Run 5 (after the installation of the additional quartz light-guides, dashed green line). One can see the effect of the additional quartz light-guides, which suppress γ quanta from PMTs. The energies of the γ lines from residual contamination are in keV.

**3.2. Data analysis**

The time-amplitude analysis, the pulse-shape discrimination between β(γ) and α particles, the pulse-shape analysis of the double pulses, and the Monte Carlo simulation of the measured energy distributions have been applied to estimate the radioactive contamination of the ZnWO$_4$ crystals.

*3.2.1. Time-amplitude analysis*

The technique of the time-amplitude analysis is described in details in [23, 24]. The arrival time and energy of each event have been used for the selection of the following fast decay chain in the $^{232}$Th family: $^{220}$Rn ($Q_\alpha$ = 6.41 MeV, $T_{1/2}$ = 55.6 s) → $^{216}$Po ($Q_\alpha$ = 6.91 MeV, $T_{1/2}$ = 0.145 s) → $^{212}$Pb. All events within 0.5 – 1.75 MeV have been used as triggers, while a time

---

particles ($E_\alpha$). Because γ quanta interact with detector by β particles, we use more convenient term "α/β ratio". The α/β ratio for ZnWO$_4$ scintillator was taken from [5].

interval 0.026 – 1.45 s (88.2% of $^{216}$Po decays) and the same energy window have been set for the second events. Sixty events of the fast chain $^{224}$Ra → $^{220}$Rn → $^{216}$Po → $^{212}$Pb were found in the data of Run 5. Taking into account the efficiency of the events selection in the time interval, one can calculate the activities of $^{228}$Th in the ZnWO$_4$ crystal ZWO-4 as 18(2) µBq/kg. The search for the fast decay chain from the $^{227}$Ac ($^{235}$U) family has also been performed in a similar way. Twelve events $^{219}$Rn ($Q_\alpha$ = 6.95 MeV, $T_{1/2}$ = 3.96 s) → $^{215}$Po ($Q_\alpha$ = 7.53 MeV, $T_{1/2}$ = 1.78 ms) → $^{211}$Pb have been selected from the data of Run 5. Thus, the activity of $^{227}$Ac in the crystal ZWO-4 has been calculated as 11(3) µBq/kg. The activities of $^{228}$Th and $^{227}$Ac in the ZnWO$_4$ crystal scintillators obtained by the time-amplitude analysis are presented in Table 3.

*3.2.2. Pulse-shape discrimination (PSD) between β(γ) and α particles*

As demonstrated in [5], the dependence of the pulse shapes on the type of irradiation in the ZnWO$_4$ scintillator allows one to discriminate γ(β) events from those induced by α particles. The optimal filter method proposed by E. Gatti and F. De Martini in 1962 [25] has been applied for this purpose.

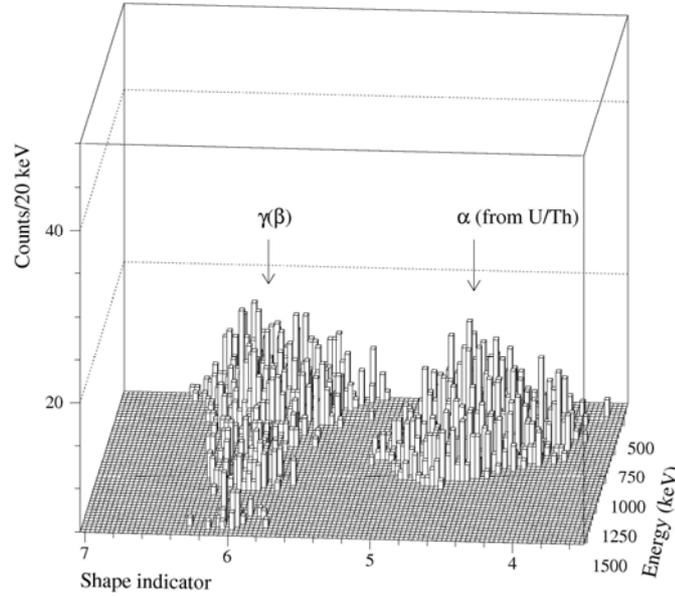

Figure 4. The three-dimensional distribution of the background events accumulated during an exposure of 4305 h with the ZnWO$_4$ crystal (ZWO-4) versus the energy and the shape indicator (see text). The population of the α events belonging to the internal contamination by U/Th is clearly separated from the γ(β) background. The energy scale refers to the calibrations with γ sources.

For each signal $f(t)$, the numerical characteristic of its shape (shape indicator, *SI*) is defined as: $SI = \sum_k f(t_k) \cdot P(t_k) / \sum_k f(t_k)$, where the sum is over the time channels $k$, starting from the origin of signal and up to 50 µs, and $f(t_k)$ is the digitized amplitude (at the time $t_k$) of a given signal. The weight function $P(t)$ is defined as: $P(t) = \{\overline{f_\alpha}(t) - \overline{f_\gamma}(t)\} / \{\overline{f_\alpha}(t) + \overline{f_\gamma}(t)\}$, where $\overline{f_\alpha}(t)$ and $\overline{f_\gamma}(t)$ are the reference pulse shapes for α particles and γ quanta, respectively. The reference shapes have been obtained by summing up a few thousands both γ and α events. For illustration, the results of the pulse-shape analysis of the data accumulated in Run 5 (for the events with the energy above 450 keV) are presented in Fig. 4 as a three-dimensional distribution of the background events versus the energy and shape indicator. One can see clearly separated the populations of α (internal contamination of the ZnWO$_4$ crystal by U/Th) and γ(β) events (mainly due to the external γ background).

The energy distributions of the γ(β) and α events selected with the help of the pulse-shape discrimination from the data of the Run 5 are shown in Fig. 5. As it was demonstrated in [5], the energy resolution for α particles is somewhat worse than that for γ quanta due to the dependence of the α/β ratio on the direction of the α particles relatively to the ZnWO$_4$ crystal axes; this makes difficult the interpretation of the α spectra. Therefore, we set only limits on the activities of the α active U/Th daughters in the ZnWO$_4$ crystals. The total α activity in the ZnWO$_4$ sample, associated with the peak in the energy interval 0.4 – 1.5 MeV, is 2.3(2) mBq/kg.

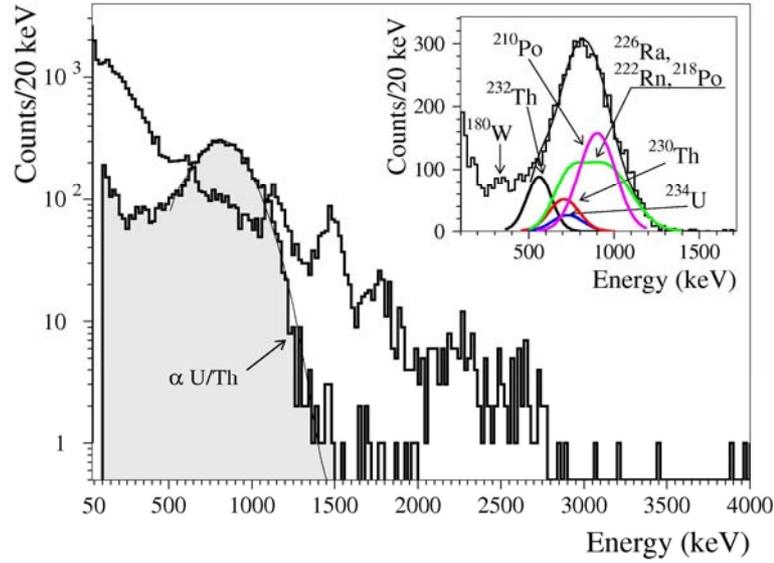

Figure 5. (Color online) The energy distributions of the β particles (γ quanta, solid histogram) and α particles (filled histogram) selected by applying the PSD procedure to the raw data measured with the ZWO-4 scintillator during 4305 h in the low background set-up. In the inset, the α spectrum is depicted together with the model, which includes α decays from $^{238}$U and $^{232}$Th families.

Besides, a small α peak with the energy in γ scale ≈ 325 keV is visible in the spectrum of the ZnWO$_4$ detector (see Fig. 5). As it was demonstrated in the measurements with a low background $^{116}$CdWO$_4$ [26] and a CaWO$_4$ [27] scintillation detectors, an α peak from the decay of $^{180}$W with half-life $T_{1/2}^{\alpha} \approx 10^{18}$ yr is expected at this energy (see subsection 4.2 for details).

A search for the fast decays $^{214}$Bi ($Q_\beta$ = 3.27 MeV, $T_{1/2}$ = 19.9 m) → $^{214}$Po ($Q_\alpha$ = 7.83 MeV, $T_{1/2}$ = 164 μs) → $^{210}$Pb (in equilibrium with $^{226}$Ra from the $^{238}$U chain) has been performed with the help of the pulse-shape analysis of the double pulses. The technique of the analysis is described in [28, 29]. The analysis gives an estimation of the activity of $^{226}$Ra in the ZWO-4 crystal as 25(6) μBq/kg. Data based on this approach for all the ZnWO$_4$ samples are presented in Table 3.

*3.2.3. Simulation of β(γ) background*

There are several β active nuclides ($^{40}$K, $^{60}$Co, $^{87}$Rb, $^{90}$Sr–$^{90}$Y, $^{137}$Cs, $^{207}$Bi, some U/Th daughters such as $^{234m}$Pa, $^{228}$Ac, $^{208}$Tl, $^{214}$Bi, $^{214}$Pb, $^{212}$Bi, $^{212}$Pb, $^{210}$Bi, $^{210}$Pb) which could produce background in the ZnWO$_4$ detectors. The radioactive contamination of the PMTs also contributes to the background. In Ref. [18] we have taken into account external γ background caused only by radioactive contamination of PMTs. As one can see from Fig. 3 (Bottom), some suppression of the background was reached after the installation of the additional quartz light-guides. However, the γ background of the detector is still considerable. Moreover, we have decided to take also into account a contribution from the copper box. The energy distributions of the possible background components were simulated with the help of the GEANT4 package [30]. The initial kinematics of the particles emitted in the decay of nuclei was given by the event generator DECAY0 [31].

The measured background spectra have been fitted by the model built from the simulated distributions. The activities of the U/Th daughters in the crystals have been restricted taking into account the results of the time-amplitude and pulse-shape analyses. The initial values of the $^{40}$K, $^{232}$Th and $^{238}$U activities inside the PMTs have been taken from [32], while activities inside the copper box have been assumed to be equal to the estimation obtained in Ref. [33].

The peak in the spectra of Runs 2, 3, 4, and 5 at the energy (1133 ± 8) keV can't be explained by contribution from external γ rays (in particular, the 1120 keV γ line of $^{214}$Bi is not enough intense to provide the whole peak area). Thus, to explain the peak we suppose the presence of $^{65}$Zn ($T_{1/2}$ = 244.26 d, $Q_{EC}$ = 1351.9 keV [34]) in the crystals[4]. The $^{65}$Zn can be produced from $^{64}$Zn by thermal neutrons (the cross section of $^{64}$Zn to thermal neutrons is 0.76 barn [34]). The fit of the background spectra gives the activity 0.5 – 0.8 mBq/kg in the ZnWO$_4$ samples ZWO-2, ZWO-3 and ZWO-4, while only the limit ≤ 0.8 mBq/kg has been obtained for activity of $^{65}$Zn in the ZWO-1 crystal. It is worth noting that the expected activity of $^{65}$Zn in a steady condition and without considering any neutron shielding deep underground at the LNGS is lower than 1 μBq/kg. There are no other clear peculiarities in the spectra which could be ascribed to the internal trace contamination by radioactive nuclides. Therefore, we can obtain only limits on the activities of the β active radionuclides and U/Th daughters. For instance, a fit of the energy spectrum of the β(γ) events (identified by the PSD) collected by the detector ZWO-4 (Run 5) in the energy region 0.1 – 2.9 MeV and the main components of the background are shown in Fig.6.

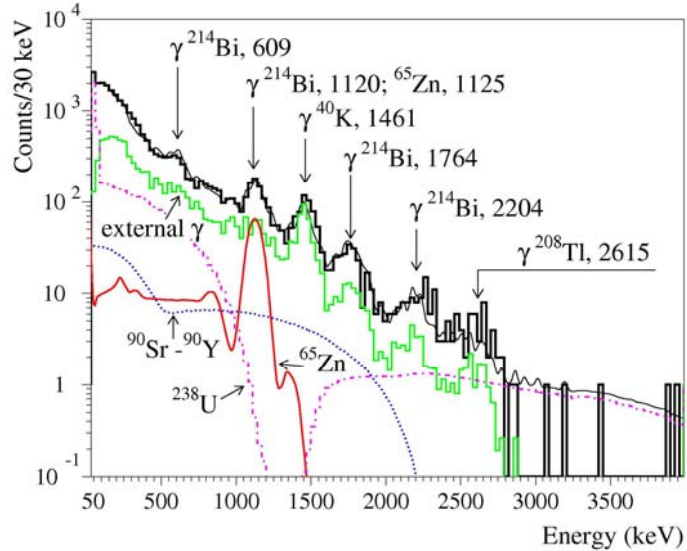

Figure 6. (Color online) Energy distribution of the β(γ) events (identified by the PSD) accumulated in the low background set-up with the ZWO-4 crystal scintillator over 4305 h (Run 5) together with the model of the background. The main components of the background are shown: spectra of internal $^{65}$Zn, $^{90}$Sr-$^{90}$Y, daughters of $^{238}$U, and the contribution from the external γ quanta from PMTs and Cu box in these experimental conditions.

The summary of the radioactive contamination of the ZnWO$_4$ crystal scintillators (or limits on their activities) is given in Table 3.

---

[4] $^{65}$Zn decays in 50.6% to the second excited level 5/2$^-$ of $^{65}$Cu [27]. Taking into account the binding energy of electrons on the K shell of nickel atoms $E_K$ ≈ 9 keV and the energy of 5/2$^-$ level $E_{ex}$ ≈ 1115.6 keV, one expects to detect a peak at the energy ≈ 1125 keV.

Table 3. Radioactive contamination of ZnWO$_4$ scintillators determined by different methods.

| Chain | Nuclide | Activity (mBq/kg) | | | | |
|---|---|---|---|---|---|---|
| | | ZWO-1 | ZWO-2 | part of ZWO-2 | ZWO-3 | ZWO-4 |
| $^{232}$Th | $^{232}$Th | ≤ 0.11 [a] | ≤ 0.1 [a] | – | ≤ 0.03 [a] | ≤ 0.25 [a] |
| | $^{228}$Ra | ≤ 0.2 [b] | ≤ 0.05 [b] | ≤ 3.4 [d] | ≤ 0.02 [b] | ≤ 0.1 [b] |
| | $^{228}$Th | 0.005(3) [c] | 0.002(1) [c] | ≤ 8.3 [d] | 0.002(2) [c] | 0.018(2) [c] |
| $^{235}$U | $^{227}$Ac | ≤ 0.007 [c] | ≤ 0.003 [c] | – | ≤ 0.01 [c] | 0.011(3) [c] |
| $^{238}$U | $^{238}$U + $^{234}$U | ≤ 0.1 [a] | ≤ 0.08 [a] | – | ≤ 0.2 [a] | ≤ 0.12 [a] |
| | $^{230}$Th | ≤ 0.13 [a] | ≤ 0.07 [a] | – | ≤ 0.15 [a] | ≤ 0.16 [a] |
| | $^{226}$Ra | ≤ 0.006 [a] | 0.002(1) [a] | ≤ 5.7 [d] | 0.021(15) [a] | 0.025(6) [a] |
| | $^{210}$Po | ≤ 0.2 [a] | ≤ 0.06 [a] | – | ≤ 0.01 [a] | ≤ 0.64 [a] |
| Total α activity | | 0.38(5) [a] | 0.18(3) [a] | – | 0.47(7) [a] | 2.3(2) [a] |
| | $^{40}$K | ≤ 1 [b] | ≤ 0.4 [b] | ≤ 24 [d] | ≤ 0.1 [b] | ≤ 0.02 [b] |
| | $^{60}$Co | ≤ 0.05 [b] | ≤ 0.1 [b] | ≤ 2.5 [d] | ≤ 0.03 [b] | ≤ 0.03 [b] |
| | $^{65}$Zn | ≤ 0.8 [b] | 0.5(1) [b] | ≤ 1.5 [d] | 0.8(2) [b] | 0.7(2) [b] |
| | $^{87}$Rb | ≤ 2.6 [b] | ≤ 2.3 [b] | – | ≤ 4.0 [b] | ≤ 4.2 [b] |
| | $^{90}$Sr-$^{90}$Y | ≤ 0.6 [b] | ≤ 0.4 [b] | – | ≤ 0.1 [b] | ≤ 0.1 [b] |
| | $^{137}$Cs | ≤ 0.3 [b] | ≤ 0.05 [b] | ≤ 1.7 [d] | ≤ 0.5 [b] | ≤ 1.3 [b] |
| | $^{147}$Sm | ≤ 0.01 [a] | ≤ 0.01 [a] | – | ≤ 0.01 [a] | ≤ 0.05 [a] |
| | $^{207}$Bi | ≤ 0.2 [b] | ≤ 0.2 [b] | ≤ 1.4 [d] | ≤ 0.4 [b] | ≤ 0.2 [b] |

[a] Pulse-shape discrimination (see subsection 3.2.2);
[b] Fit of background spectra (see subsection 3.2.3);
[c] Time-amplitude analysis (see subsection 3.2.1);
[d] HP Ge γ spectrometry (see subsection 3.3).

### 3.3. Ultra-low background HP Ge γ spectrometry

A crystal part (∅44 × 41 mm, mass 514 g) cut from the ZWO-2 sample has been measured with the help of the ultra-low background 244 cm$^3$ HP Ge γ spectrometer GeBer at the LNGS. The sample was placed directly on the end-cap of the detector.

To reduce the external background, the detector is shielded by layers of low radioactive copper (≈ 10 cm), lead (≈ 20 cm) and polyethylene (≈ 10 cm). The set-up has been continuously flushed by high purity nitrogen (stored deep underground for a long time) to avoid presence of residual environmental radon. The ZnWO$_4$ sample was measured over 1058 h; the background of the detector was accumulated during 3047 h.

The energy distribution accumulated with the sample is shown in Fig. 7 in the energy region 20 – 2900 keV together with the background data. The energy distribution measured with the sample practically coincides with the background; thus, only limits for possible radioactive contaminations have been derived from the data (see Table 3). The detection efficiency of the HP Ge detector was calculated using the GEANT4 code [30]. It should be stressed that there are no peaks in the spectrum taken with the ZnWO$_4$ crystal sample which can be ascribed to the decay of internal $^{65}$Zn; therefore only limit on its activity in the ZWO-2 scintillator was set on the level of ≤ 1.5 mBq/kg. However, the sensitivity of the measurement is not high enough to observe the activity of this nuclide 0.5(1) mBq/kg detected by the scintillation method.

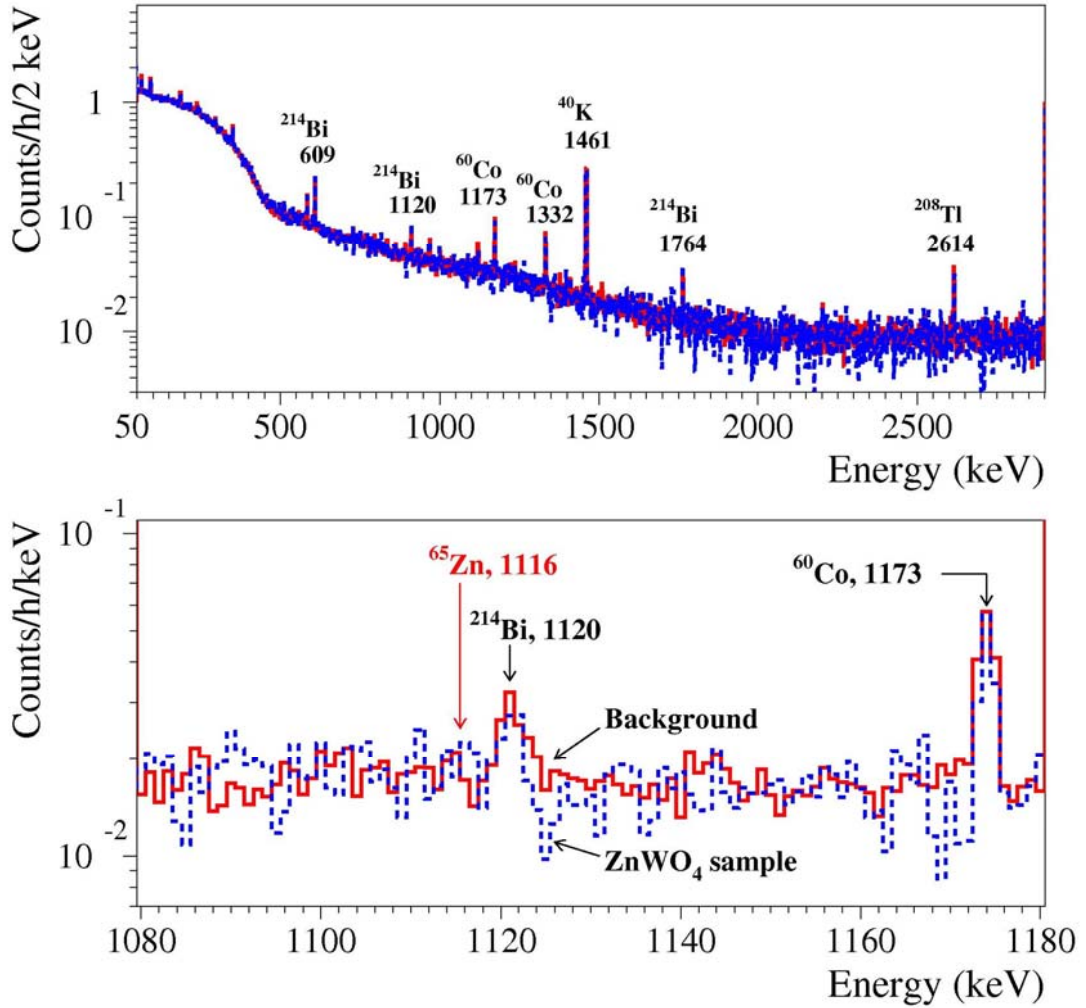

Figure 7. (Color online) The energy distribution measured with a part of ZWO-2 crystal (514 g) by the HP Ge GeBer detector (244 cm$^3$) over 1058 h (blue) in comparison with the background (red) measured during 3047 h (Top). The energy distribution measured with the ZnWO$_4$ sample practically coincides with the background. The part of the energy spectrum in the vicinity of the 1116 keV peak expected for the decay of $^{65}$Zn to the excited level of $^{65}$Cu is shown (Bottom). The spectra are normalized on the time of measurements; the energies of the γ lines are in keV.

### 3.4. Radioactive contamination of the ceramics used in the set-ups employed in the crystals growth

Measurements of the radioactive contamination of the ceramic materials used in the set-ups for the ZnWO$_4$ crystals growth have been performed in the Institute for Nuclear Research (Kyiv, Ukraine) with the help of a low background HP Ge detector (73 cm$^3$, CANBERRA, model GR 1519). The samples of Ceramics-1, 2 and 3 were provided by the ISMA (Kharkiv, Ukraine), the samples Ceramics-4 and 5 were sent by the NIIC (Novosibirsk, Russia). The measurements have been carried out over a few days for each sample. As one could expect, the ceramics (especially Ceramics-1 and Ceramics-2 used in the ISMA) are rather radioactive materials (see Table 4).

Table 4. Radioactive contamination of ceramics, used in the set-ups where the studied $ZnWO_4$ crystals were grown, measured by low background HP Ge γ spectrometer.

| Chain | Nuclide | Activity (Bq/kg) in samples. Mass of samples and time of measurements | | | | |
|---|---|---|---|---|---|---|
| | | Ceramics-1 [a] ISMA 200 g 74.0 h | Ceramics-2 [b] ISMA 237 g 49.0 h | Ceramics-3 [c] ISMA 118.6 g 141.1 h | Ceramics-4 [d] NIIC 61 g 46.4 h | Ceramics-5 [e] NIIC 46.4 g 93.6 h |
| $^{232}$Th | $^{214}$Pb | 50(2) | 22(2) | ≤ 0.9 | ≤ 0.4 | ≤ 7 |
| | $^{214}$Bi | 36(1) | 16(1) | ≤ 0.8 | ≤ 2.4 | 2(1) |
| $^{238}$U | $^{228}$Ac | 42(1) | 18(1) | ≤ 2 | ≤ 3 | ≤ 3 |
| | $^{207}$Bi | ≤ 0.2 | ≤ 0.6 | ≤ 0.7 | ≤ 1 | ≤ 1 |
| | $^{137}$Cs | ≤ 1 | ≤ 1 | ≤ 1 | ≤ 1 | 2(1) |
| | $^{40}$K | 215(7) | 102(6) | ≤ 8 | 21(9) | ≤ 9 |
| | $^{60}$Co | ≤ 0.2 | ≤ 0.1 | ≤ 0.3 | ≤ 0.3 | ≤ 0.4 |

[a] Corundum ceramics, 98% of $Al_2O_3$, Ukrainian Research Institute for Refractories, Ukraine;
[b] Mullite corundum, consists of $SiO_2$ and $Al_2O_3$, (72 − 90)% of $Al_2O_3$, factory "POLIKOR" Russia;
[c] Ceramic grit, Ukrainian Research Institute for Refractories, Ukraine;
[d] Kaolin wool, MKRR-130 GOST 23619-79, SC Sukholuzhskie ogneupory, Sverdlov region, Suhoi Log, Russia;
[e] Parts of ceramic bricks, Heat-insulating material TKT TU 81-04-437-76, Krasnogorodskoye experimental paper mill, Krasnoe Selo, First May Str. 1, 2, Russia.

## 4. RESULTS AND DISCUSSION

### 4.1. Radiopurity of $ZnWO_4$ scintillators

The radioactive contamination of the $ZnWO_4$ crystal scintillators is summarized in Table 5 where the data for other tungstates ($CaWO_4$ [27, 35], $CdWO_4$ [36, 37]), and for specially developed low background NaI(Tl) [38] are given for comparison. The radioactive contamination of all the samples of $ZnWO_4$ crystals, studied in the present work, is at the level of 0.002 - 0.8 mBq/kg (depending on the source), which approaches that of $CdWO_4$ and of NaI(Tl), and is considerably lower than that of $CaWO_4$. In future creation of $ZnWO_4$ crystal scintillators new careful measurements of the initial materials have to be carried out since the radioactive contaminations of the initial compounds, used in the crystal growth, give considerable contribution to the radioactivity of the crystal scintillators.

From the comparison of the data for the ZWO-2 and ZWO-3 samples one can conclude that the re-crystallization does not improve the radiopurity level of the $ZnWO_4$ crystal. Anyhow, further investigations of the effect of the re-crystallization on the radioactive contamination of the $ZnWO_4$ crystal scintillators could also be carried out taking into account possible inhomogeneity of U/Th traces distribution in the crystal volume.

Table 5. Comparison of radioactive contamination of some $ZnWO_4$, $CaWO_4$, $CdWO_4$, and NaI(Tl) crystal scintillators.

| Source | Activity (mBq/kg) | | | | |
|---|---|---|---|---|---|
| | $ZnWO_4$ Present study | $ZnWO_4$ [5, 36, 16, 9] | $CaWO_4$ [27, 35] | $CdWO_4$ [36, 37] | NaI(Tl) [38] |
| $^{228}$Th | 0.002 – 0.018 | ≤ (0.1 – 0.2) | 0.6 | ≤ 0.004 – 0.04 | 0.002 |
| $^{226}$Ra | 0.002 – 0.025 | ≤ (0.16 – 0.4) | 6 | ≤ (0.004 – 0.04) | 0.009 |
| Total α activity | 0.2 – 2 | ≤ 2 – 20 | 20 – 400 | 0.3 – 2 | 0.08 |
| $^{40}$K | ≤ (0.01 – 1) | ≤ 12 | ≤ 12 | 0.3 | < 0.6 |
| $^{65}$Zn | 0.5 – 1 | – | – | – | – |
| $^{90}$Sr | ≤ (0.01 – 0.6) | ≤ (1.2 – 15) | ≤ 70 | ≤ 0.2 | – |
| $^{113}$Cd | – | – | – | 558 | – |
| $^{137}$Cs | ≤ (0.05 – 1.2) | ≤ (2.5 – 20) | ≤ 0.8 | ≤ 0.3 – 0.4 | – |
| $^{147}$Sm | ≤ (0.01 – 0.05) | ≤ (1.8 – 5) | 0.5 | ≤ (0.01 – 0.04) | – |
| $^{180}$W | 0.04 | – | 0.05 | 0.04 | – |

The ceramics samples used in the set-ups to grow the studied $ZnWO_4$ crystal scintillators are contaminated by U/Th and $^{40}$K on the level which exceeds the radioactive contamination of the studied $ZnWO_4$ crystal scintillators almost four orders of magnitude. In a naive consideration of growing process, a crystal ingot cannot be polluted by vapors from other (colder) parts of the set-up. Indeed, assuming that the crystal has higher temperature than that of the ceramics[5], one might conclude that the radioactivity of the ceramics is not so dangerous (vapors of radioactive elements cannot condense on hotter crystal). However, in a real set-up one cannot exclude pollution of crystal on the level of $10^{-4}$ from the ceramics contamination through dust spread, chipping of micro-size ceramics particles, etc. Thus, the study of the possible effects of the radioactive contamination of the details of the set-ups used in the crystals growth on the radiopurity of the crystal scintillators should be a subject of additional extensive investigations.

The Czochralski method has been used in the $ZnWO_4$ crystals growth, and the crucible — which is always the hottest part — was made of platinum. Recent measurements of samples of this material at the LNGS by ultra-low background HP Ge γ spectrometry have shown comparatively low level of radioactive contamination (preliminary, in those latter samples of platinum the activity of $^{214}$Bi and $^{228}$Th is on the level of a few mBq/kg [39], which is comparable to the radioactive contamination of the $ZnWO_4$ crystals). We remind that special platinum crucible suitably cleaned and conditioned was — after several tests — selected as the best one to grow ultra-low background NaI(Tl) by Kyropoulos method [38].

Finally, $ZnWO_4$ crystals having higher radiopurity could be expected in future realizations by careful selection and purification of the initial materials. In particular, one could expect that vacuum distillation and filtering could be very promising approaches [40, 41] to obtain high purity zinc, while zone melting could be used for additional purification of tungsten. Obviously all the more accurate techniques for the screening, purification and protection from environment of all the materials in every stage should be applied.

### 4.2. Alpha-decay of $^{180}$W

An indication of the alpha decay of $^{180}$W (the expected energy of alpha particles is 2460(5) keV [42], isotopic abundance of $^{180}$W is δ = 0.12% [43]), with a half-life $T_{1/2}$ ~ $10^{18}$ yr was obtained in measurements with low background $CdWO_4$ and $CaWO_4$ crystal scintillators (see Table 6).

---

[5] In the high gradient Czochralski method with high frequency hitting a crystal is hotter in comparison to ceramic details of growth set-up. Conversely, in the low-thermal gradient technique crystal is colder in comparison to the heater made from high-resistance wire wound on a ceramic.

Table 6. Summary of experiments on $^{180}W$ α decay.

| Experimental | $T_{1/2}$ (yr) | Year, Reference |
|---|---|---|
| Three $^{116}CdWO_4$ crystal scintillators (330 g, enriched in $^{116}Cd$ to 83%), Solotvina Underground Laboratory (1000 m w.e.), 2975 h | $(1.1^{+0.8}_{-0.4} \pm 0.3) \times 10^{18}$ | 2003 [26] |
| $CaWO_4$ crystal scintillator as a bolometer (10 mK), Gran Sasso National Laboratory (3600 m w.e.), 6701 h | $(1.8 \pm 0.2) \times 10^{18}$ | 2004 [35] |
| $CaWO_4$ crystal scintillator (189 g), Solotvina Underground Laboratory (1000 m w.e.), 1734 h | $(1.0^{+0.7}_{-0.3}) \times 10^{18}$ | 2005 [27] |
| Present experiment | $(1.3^{+0.6}_{-0.5}) \times 10^{18}$ | 2010 |

A peculiarity in the α spectrum of the $ZnWO_4$ detectors at the energy of 325(11) keV (see Fig. 8) corresponds to the α particle energy of 2358(80) keV. These alpha events can be ascribed to the α decay of $^{180}W$. A 100 g $ZnWO_4$ crystal contains $2.31 \times 10^{20}$ nuclei of $^{180}W$. The area of the peak is $(204 \pm 75)$ counts, which gives, taking into account the 49.7% efficiency of the pulse-shape selection technique, $(4.1 \pm 1.5) \times 10^2$ decays of $^{180}W$. Thus, one can derive the half-life of $^{180}W$ relatively to α decay as $T_{1/2} = (1.3^{+0.6}_{-0.5}) \times 10^{18}$ yr. This result is in agreement with all the data published earlier.

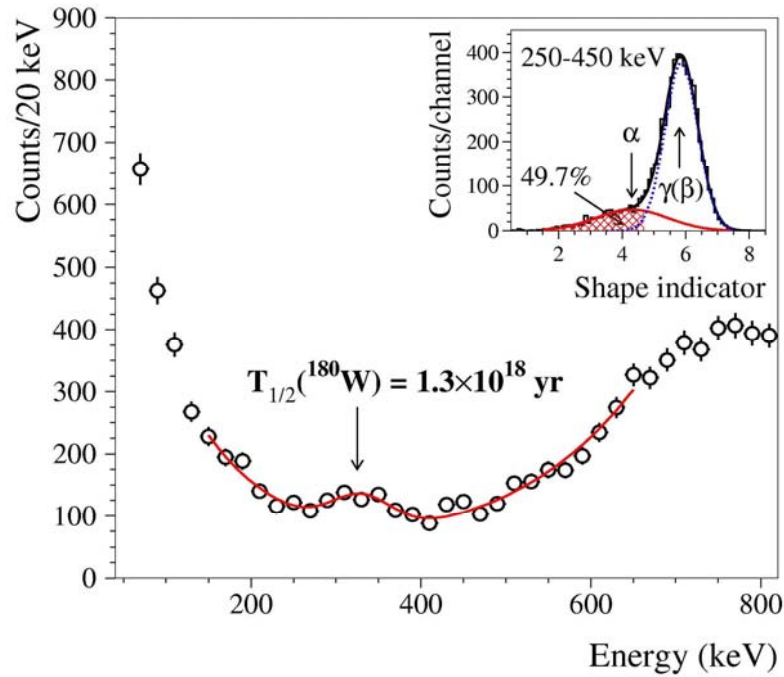

Figure 8. (Color online) Fragment of the α spectrum selected by the pulse-shape discrimination from the data measured with the $ZnWO_4$ detectors in Runs 1 – 5 over 3197 kg × h together with the fitting curve (solid line). The α peak of $^{180}W$ with the area of 204 counts corresponds to the half-life $1.3 \times 10^{18}$ yr. In the inset, the shape indicator distributions measured with the ZWO-4 scintillator during 4305 h in the energy range 250 – 450 keV. Fitting curves correspond to β particles (γ quanta, dotted blue line) and α particles (solid red line).

## 5. CONCLUSIONS

The residual contamination in the measured ZnWO$_4$ cystal scintillators (total α activity) is at the level of (0.2 – 2) mBq/kg. Particular contamination, i.e. associated with the elemental composition of the ZnWO$_4$, was observed: the β active $^{65}$Zn (probably due to cosmogenic or/and neutron activation at sea level) and α active $^{180}$W (rare α decay with $T_{1/2}$ ~ 10$^{18}$ yr); the $^{65}$Zn radioisotope is rather dangerous for the experiments on rare processes. No improvement in the ZnWO$_4$ crystal radiopurity has been found after re-crystallization; further investigation is foreseen.

An α peak, which can be ascribed to α activity of $^{180}$W with half-life $T_{1/2} = (1.3^{+0.6}_{-0.5}) \times 10^{18}$ yr, has been observed in the energy distribution (exposure 3197 kg × h); this result is in agreement with other experiments.

The radioactive contaminations of the ceramic materials used in the set-ups employed in the crystals growth exceeds three-four orders of magnitude the radioactive contamination of the ZnWO$_4$ crystals; thus, future R&Ds are needed to further quantitatively investigate the effect of the details of the growing process on the reachable radiopurity level of the ZnWO$_4$ crystals.

**Acknowledgements**

The group from the Institute for Nuclear Research (Kyiv, Ukraine) was supported in part by the Project "Kosmomikrofizyka-2" (Astroparticle Physics) of the National Academy of Sciences of Ukraine. D.V. Poda, R.B. Podviyanuk, and O.G. Polischuk were also partly supported by the Grant for Young Scientists of the National Academy of Science of Ukraine (Reg. No. 0109U007070).